# A brief introduction to giant magnetoresistance

Liu Chang[1], Min Wang[1], Lei Liu[1], Siwei Luo[2], Pan Xiao[1*],

[1]Hycorelle Co., Ltd., Beijing, China

[2]Ohio State University, Columbus, OH 43210

*corresponding author, arm0728@live.cn

**Abstract:** Giant magnetoresistance (GMR) is a quantum mechanical magnetoresistance effect observed in thin film structures composed of alternating ferromagnetic and nonmagnetic layers. The effect manifests itself as a significant decrease (typically 10–80%) in electrical resistance in the presence of a magnetic field. The effect is exploited commercially by manufacturers of hard disk drives. The 2007 Nobel Prize in physics was awarded to Albert Fert and Peter Grünberg for the discovery of GMR.

**Keyword:** GMR, spin valve, magnetization, multilayer, hard disk

## 0 Introduction

The "giant magnetoresistive" (GMR) effect was discovered in the late 1980s by two European scientists working independently: Peter Gruenberg of the KFA research institute in Julich, Germany, and Albert Fert of the University of Paris-Sud . They saw very large resistance changes (Fig.1) -- 6 percent and 50 percent, respectively -- in materials comprised of alternating very thin layers of various metallic elements. This discovery took the scientific community by surprise; physicists did not widely believe that such an effect was physically possible. These experiments were performed at low temperatures and in the presence of very high magnetic fields and used laboriously grown materials that cannot be mass-produced, but the magnitude of this discovery sent scientists around the world on a mission to see how they might be able to harness the power of the Giant Magneto resistive effect [1-3].

Like other magnetoresistive effects, GMR is the change in electrical resistance in response to an applied magnetic field. Transition metals are extensively studies in many fields [4-26]. Their magnetoresistive effects are also been given significant interests. It was discovered that the application of a magnetic field to a Fe/Cr multilayer resulted in a significant reduction of the electrical resistance of the multilayer. 1. This effect was found to be much larger than either ordinary or anisotropic magnetoresistance and was, therefore, called "giant magnetoresistance" or GMR. A similar, though diminished effect was simultaneously discovered in Fe/Cr/Fe trilayers.2. As was shown later, high magnetoresistance values can also be obtained in other magnetic multilayers, such as Co/Cu. The change in the resistance of the multilayer arises when the applied field aligns the magnetic moments of the successive ferromagnetic layers, as is illustrated schematically in Fig.2. In the absence of the magnetic field the magnetizations of the ferromagnetic layers are antiparallel. Applying the magnetic field, which aligns the magnetic moments and saturates the magnetization of the multilayer, leads to a drop in the electrical resistance of the multilayer [1-3].

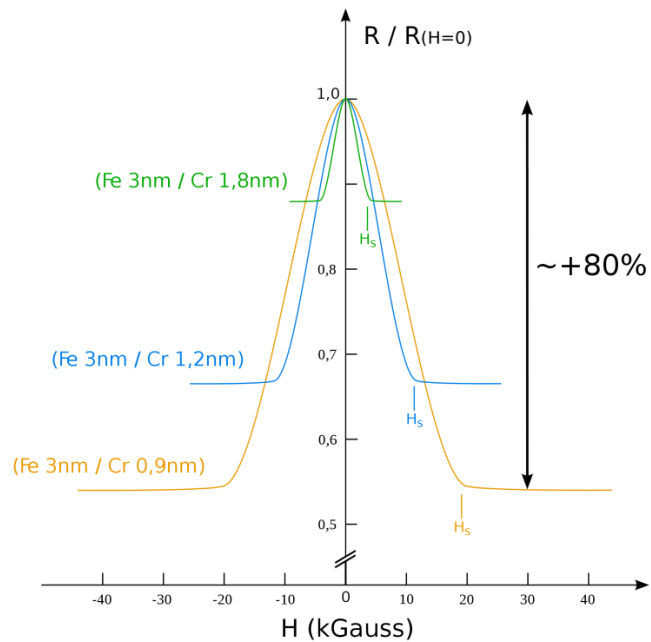

Figure 1. The founding results of Albert Fert and Peter Grünberg (adapted from [1])

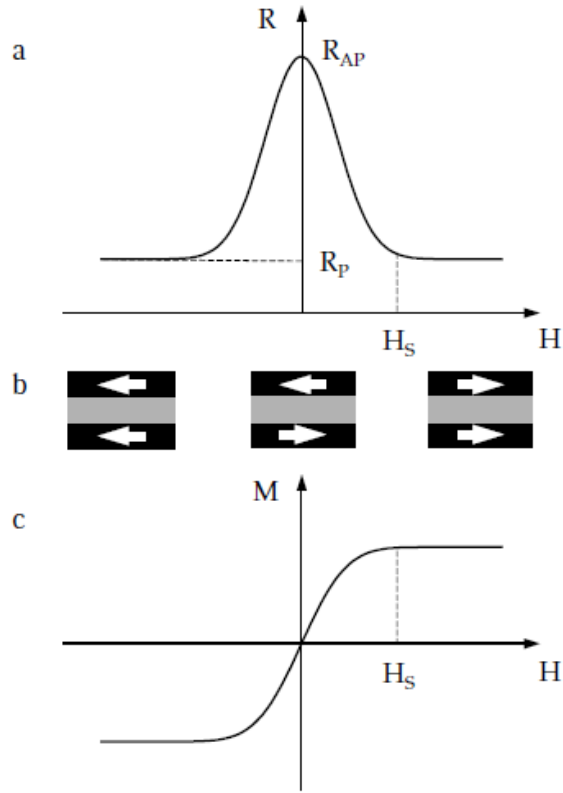

Figure.2 Schematic representation of the GMR effect. (a): Change in the resistance of the magnetic multilayer as a function of applied magnetic field. (b): The magnetization configurations (indicated by the arrows) of the multilayer (trilayer) at various magnetic fields: the magnetizations are aligned antiparallel at zero field; the magnetizations are aligned parallel when the external magnetic field H is larger than the saturation field HS. (c): The magnetization curve for the multilayer. (adapted from [3])

## 1  Origin of GMR

GMR can be qualitatively understood using the Mott model, which was introduced as early as 1936 to explain the sudden increase in resistivity of ferromagnetic metals as they are heated above the Curie temperature. There are two main points proposed by Mott [1-3]:

First, the electrical conductivity in metals can be described in terms of two largely independent conducting channels, corresponding to the up-spin and down-spin electrons, which are distinguished according to the projection of their spins along the quantization axis. The probability of spin-flip scattering processes in metals is normally small as compared to the probability of the scattering processes in which the spin is conserved. This means that the up-spin and down-spin electrons do not mix over long distances and, therefore, the electrical conduction occurs in parallel for the two spin channels [1-3].

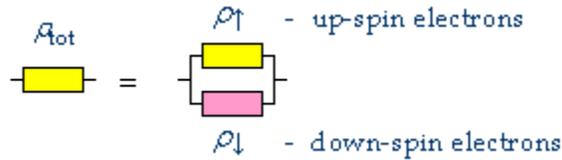

Second, in ferromagnetic metals the scattering rates of the up-spin and down-spin electrons are quite different, whatever the nature of the scattering centers is. The band structure in a ferromagnet is exchange-split, so that the density of states is not the same for up-spin and down-spin electrons at the Fermi energy. Scattering rates are proportional to the density of states, so the scattering rates and therefore resistivities are different for electrons of different spin.

$$\rho_\uparrow \neq \rho_\downarrow$$

Using Mott's arguments it is straightforward to explain GMR. We assume that the scattering is strong for electrons with spin antiparallel to the magnetization direction, and is weak for electrons with spin parallel to the magnetization direction. This is supposed to reflect the asymmetry in the density of states at the Fermi level, in accordance with Mott's second argument. For the parallel-aligned magnetic layers, the up-spin electrons pass through the structure almost without scattering, because their spin is parallel to the magnetization of the layers. On the contrary, the down-spin electrons are scattered strongly within both ferromagnetic layers, because their spin is antiparallel to the magnetization of the layers. Since conduction occurs in parallel for the two spin channels, the total resistivity of the multilayer is determined mainly by the highly-conductive up-spin electrons and appears to be low. For the antiparallel-aligned multilayer, both the up-spin and down-spin electrons are scattered strongly within one of the ferromagnetic layers, because within the one of the layers the spin is antiparallel to the magnetization direction. Therefore, in this case the total resistivity of the multilayer is high.

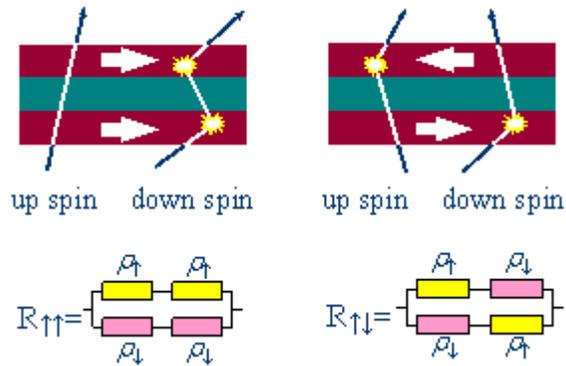

## 2 Types of GMR

### 2.1 Multilayer GMR

In multilayer GMR two or more ferromagnetic layers are separated by a very thin (about 1 nm) non-ferromagnetic spacer (e.g. Fe/Cr/Fe). At certain thicknesses the RKKY coupling between adjacent ferromagnetic layers becomes antiferromagnetic, making it energetically preferable for the magnetizations of adjacent layers to align in anti-parallel. The electrical resistance of the device is normally higher in the anti-parallel case and the difference can reach more than 10% at room temperature. The interlayer spacing in these devices typically corresponds to the second antiferromagnetic peak in the AFM-FM oscillation in the RKKY coupling [1-3].

The GMR effect was first observed in the multilayer configuration, with much early research into GMR focusing on multilayer stacks of 10 or more layers.

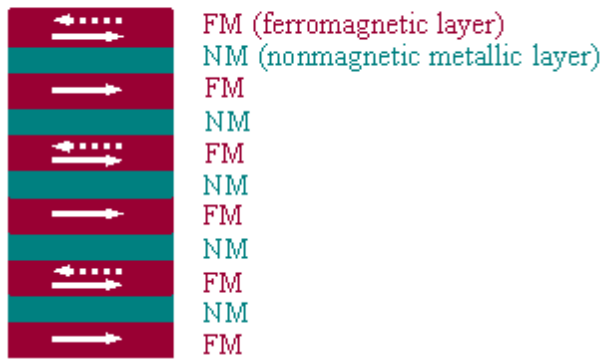

### 2.2 Spin valve GMR

In spin valve GMR two ferromagnetic layers are separated by a thin non-ferromagnetic spacer (~3 nm), but without RKKY coupling. If the coercive fields of the two ferromagnetic electrodes are different it is possible to switch them independently. Therefore, parallel and anti-parallel alignment can be achieved, and normally the resistance is again higher in the anti-parallel case. This device is sometimes also called a spin valve [1-3].

Research to improve spin valves is intensely focused on increasing the MR ratio by practical methods such as increasing the resistance between individual layers interfacial resistance, or by inserting half metallic layers into the spin valve stack. These work by increasing the distances over which an electron will retain its spin (the spin relaxation length), and by enhancing the polarization effect on electrons by the ferromagnetic layers and the interface. The magnetic properties of nanostructures (and all properties) are dominated by surface and interface effects due to the high local ratio of atoms as compared to the bulk.

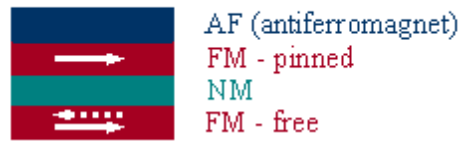

### 2.3 Pseudo-spin GMR

Pseudo-spin valve devices are very similar to the spin valve structures. The significant difference is the coercivities of the ferromagnetic layers. In a pseudo-spin valve structure a soft magnet will be used for one layer; where as a hard ferromagnet will be used for the other. This allows the applied field to flip the magnetization of one layers before the other, thus providing the same anti-ferromagnetic affect that is required for GMR devices. For pseudo-spin valve devices to work they generally require the thickness of the non-magnetic layer to be thick enough so that exchange coupling is kept to a minimum. It is imperative to prevent the interaction between the two ferromagnetic layers in order to exercise complete control over the device.

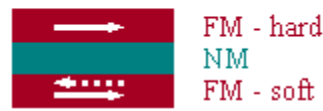

### 2.4 Granular GMR

Granular GMR is an effect that occurs in solid precipitates of a magnetic material in a non-magnetic matrix. To date, granular GMR has only been observed in matrices of copper containing cobalt granules. The reason for this is that copper and cobalt are immiscible, and so it is possible to create the solid precipitate by rapidly cooling a molten mixture of copper and cobalt. Granule sizes vary depending on the cooling rate and amount of subsequent annealing. Granular GMR materials have not been able to produce the high GMR ratios found in the multilayer counterparts [1-3].

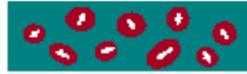

## 2.5 Comparison of four different types of GMR

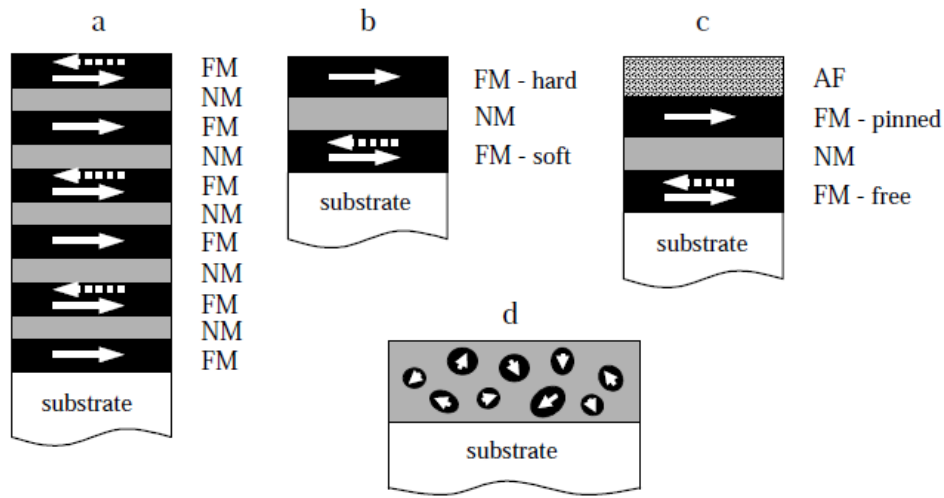

Figure.3 Various structures in which GMR can be observed: magnetic multilayer (adapted from [3])

Various structures in which GMR can be observed: magnetic multilayer (a), pseudo spin valve (b), spin valve (c) and granular thin film (d). Note that the layer thickness is of the order of a few nanometers, whereas the lateral dimensions can vary from micrometers to centimetres. In the magnetic multilayer (a) the ferromagnetic layers (FM) are separated by nonmagnetic (NM) spacer layers. Due to antiferromagnetic interlayer exchange coupling they are aligned antiparallel at zero magnetic field as is indicated by the dashed and solid arrows. At the saturation field the magnetic moments are aligned parallel (the solid arrows). In the pseudo spin valve (b) the GMR structure combines hard and soft magnetic layers. Due to different coercivities, the switching of the ferromagnetic layers occurs at different magnetic fields providing a change in the relative orientation of the magnetizations. In the spin valve (c) the top ferromagnetic layer is pinned by the attached antiferromagnetic (AF) layer. The bottom ferromagnetic layer is free to rotate by the applied magnetic field. In the granular material (d) magnetic precipitates are embedded in the non-magnetic metallic material. In the absence of the field the magnetic moments of the granules are randomly oriented. The magnetic field aligns the moments in a certain direction [1-3].

## 3 Application of GMR

The largest technological application of GMR is in the data storage industry. IBM was first to put on the market hard disks based on GMR technology, and nowadays all disk drives make use of this technology [1-3].

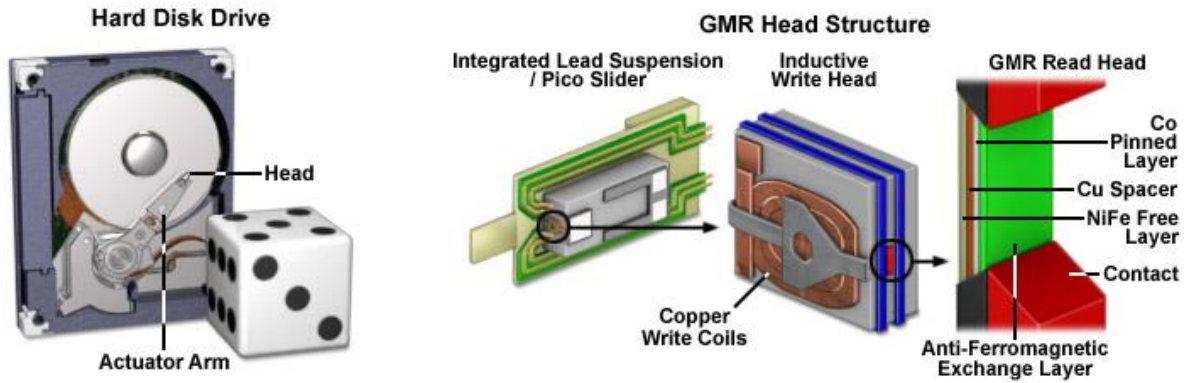

Figure 4. Structure of hard disk and GMR head (adapted from [2])

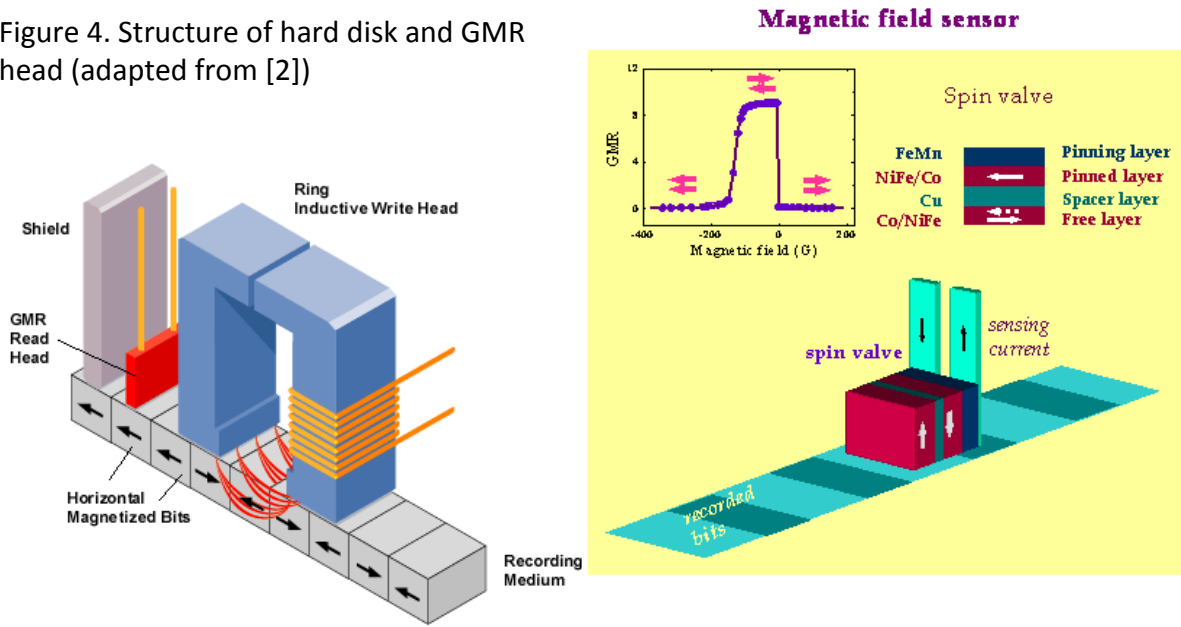

Figure 5. Illustration of reading and writing (adapted from [2])

The GMR read head sensor in a hard disk is built using a spin valve . Spin valve resistance demonstrates a steep change in the small field range close to H=0. As the magnetic bits on the hard drive pass under the read head, the magnetic alignment of the sensing layer in the spin valve changes resulting in the resistance change.

As the read head passes over the disk, the free layer shifts its magnetic orientation to match that of the bit. So sometimes the direction of the free layer's magnetic field is aligned with the field of the pinned layer (which never changes), and sometimes it is opposite. When they are aligned, most of their electrons will share the same up or down spin. As some of these electrons pass through the layers in the form of current, there will be minimal scattering. The low resistance means a current will be detected, and the computer computes a 1 bit [1-3].

When the free layer's magnetic orientation switches to opposite that of the pinned layer, there's a much different result. The electrons in the two layers have opposing spins. So as the current passes through the magnetized layers, those electrons will scatter in one or the other of them, resulting in a much weaker current and a 0 bit.

Other applications of GMR are as diverse as automotive sensors, solid-state compasses and non-volatile magnetic memories.

## 4   Conclusion

Giant magnetoresistance is the large change in electrical resistance of metallic layered systems when the magnetizations of the ferromagnetic layers are reoriented relative to one another under the application of an external magnetic field. This reorientation of the magnetic moments alters both the electronic structure and the scattering of the conduction electrons in these systems, which causes the change in the resistance. Various types of magnetic layered structures have been found which show sizable values of GMR. Highest values are obtained in magnetic multilayer structures, such as Fe/Cr and Co/Cu, which remain attractive from the point of view of studying the fundamental physics involved. The exchange-biased spin valves show a combination of properties that make these systems more useful for applications in low-field sensors, such as read heads for magnetic recording.